\documentclass[aps,a4paper,superscriptaddress,nofootinbib,showpacs,showkeys,amsfonts,amssymb,amsmath]{revtex4}

\usepackage{amssymb,latexsym}
\usepackage{amsmath, amsthm}
\usepackage{amscd}
\usepackage{times}
\usepackage{epsfig}
\usepackage{psfrag}
\usepackage{graphicx}

\begin{document}
\title{Instability of black hole formation under small pressure perturbations}

\author{Pankaj S. Joshi} \email{psj@tifr.res.in}
\affiliation{Tata Institute of Fundamental Research, Homi Bhabha road,
Colaba, Mumbai 400005, India}
\author{Daniele Malafarina} \email{daniele.malafarina@polimi.it}
\affiliation{Tata Institute of Fundamental Research, Homi Bhabha road,
Colaba, Mumbai 400005, India}

\swapnumbers

\begin{abstract}

We investigate here the spectrum of gravitational
collapse endstates when arbitrarily small
perfect fluid pressures are introduced in the classic black hole
formation scenario as described by Oppenheimer, Snyder and Datt (OSD)
\cite{OSD}.
This extends a previous result on tangential pressures \cite{MJ}
to the more physically realistic scenario of perfect fluid collapse.
The existence of classes of pressure perturbations
is shown explicitly, which has the property that injecting
any smallest pressure changes the final fate of the
dynamical collapse from a black hole to a
naked singularity. It is therefore seen that any smallest
neighborhood of the OSD model, in the space of initial
data, contains collapse evolutions that go to a naked
singularity outcome. This gives an intriguing insight
on the nature of naked singularity formation in
gravitational collapse.

\end{abstract}
\pacs{04.20.Dw,04.20.Jb,04.70 Bw}
\keywords{Gravitational collapse, black holes, naked singularity}
\maketitle

\section{Introduction}

One of the most important open
issues in the theory and astrophysical applications of
modern day black hole and gravitation physics
is that of the Cosmic Censorship Conjecture (CCC)
\cite{CCC}.
The CCC postulates that any physically
realistic gravitational
processes must not lead to the formation
of a singularity which is not covered by an horizon,
thus hiding it from external observers in the universe.
This of course includes the complete
gravitational collapse of a massive star
which, if the CCC is true, must terminate generically into
a black hole final state only.
Nevertheless in recent years, a wide variety
of gravitational collapse models have been discovered
where the dynamical evolution leads to a naked singularity
formation rather than a black hole as collapse endstate
(see e.g. \cite{NS}).

The consideration of dynamical evolution is a crucial
element of the CCC in the sense that in the original
formulation of the conjecture it is stated that singularities
arising from the dynamical collapse from a regular initial
data must be covered by an event horizon.
Many solutions of Einstein field equations are known
which present naked singularities (such as, for example,
the super-spinning Kerr solutions), nevertheless almost
none of these solutions can be obtained as the dynamically
evolved final state of some initially regular matter
configuration. For this reason, over the last decades
a great deal of work has been done to test the CCC in the
few dynamically evolving spacetimes we know. These are
typically the scenarios that describe gravitational collapse
in spherical symmetry, and some non-spherical collapse
models have also been considered
\cite{JK}.

In the following, we will consider complete gravitational
collapse of a spherical massive matter cloud that leads to the
formation of a strong shell focusing naked singularity at the center.
In such a case, the super-ultra-dense regions, or the spacetime
singularity, that forms at the end of collapse would be
visible to faraway observers in the universe, rather
than being hidden in a black hole. Therefore, the question
crucial now for the CCC is that of how stable and generic
are the naked singularities and black holes that form in
dynamical gravitational collapse of a massive
matter cloud.

The astrophysical significance of the issue,
and the importance of considering the gravitational
collapse of a matter cloud within the framework of the
general relativity theory, with reasonable physical
properties for the matter included, stems from the fact that
a star more massive than about five to eight times the
mass of the Sun, cannot stabilize to a neutron star
final state at the end of its life cycle.
It must collapse continually under the force
of its own gravity on exhausting its
internal nuclear fuel, and there are no known forces of
nature that would halt such a collapse. General relativity
predicts that such a star must then terminate into a spacetime
singularity where densities and spacetime curvatures
blow up and the physical conditions are extreme. The CCC
assumption that such a singularity is always covered within
an event horizon of gravity, is then crucial and
is at the basis of much of the modern theory and astrophysical
applications of black holes today. However, despite the past
four decades of serious efforts, we do not have as yet
available any proof or even any mathematically precise
formulation of the cosmic censorship hypothesis.

Actually from the many dynamical gravitational
collapse scenarios that have been investigated over
the past years, the typical conclusion that has
emerged is as follows: Depending on the nature
of the initial data for the matter cloud in terms
of its initial density, pressure and velocity profiles
from which the collapse evolves, there are dynamical
evolutions governed by the Einstein equations that
take the collapse to either a black hole or naked
singularity final state (see e.g.
\cite{Joshi}
and references therein).
Such a naked singularity would be a super-ultra-dense
region of extreme gravity that can communicate with
faraway observers in spacetime, and this hypothetical
astrophysical object, if realized in nature,  would
have radically different observational signatures
from its black hole counterpart.

If, however, the occurrence of naked singularities
of collapse were special or `non-generic' in some appropriately
well-defined sense, then at least the spirit of
CCC would be respected. One can impose various
sets of physical reasonability and regularity conditions
under which the collapse of
a massive star is to be dynamically evolved to examine
its final state.
What is really needed here is a
detailed investigation of the gravitational collapse
phenomena within the framework of general relativity,
which is the only path that can provide useful and adequate
insights into the final fate of collapse in terms of
either a black hole or naked singularity.
Investigating how the final black hole state
is affected, once a small, general, perfect fluid pressure
is introduced in the initial data from which
the collapse develops would provide a much better
understanding of the collapse final states.
We show here that the models ending in a naked
singularity are in fact not `special', in the sense that
they can be generally obtained from perfect fluid collapse,
where the initial data is arbitrarily close to models
leading to a black hole.

Specifically, we examine here how the evolution
of dust collapse models having a black hole final state, is altered
when an arbitrarily small perfect fluid pressure perturbation
is introduced in the matter source. We show
explicitly the existence of classes of small generic
pressure perturbations such that an injection of
a small positive (or negative) pressure in the OSD model, or in a
Tolman-Bondi-Lemaitre (TBL) inhomogeneous dust collapse
to a black hole
\cite{LTB},
leads the collapse to form a
naked singularity, rather than a black hole.
The classic OSD scenario is the basic paradigm
for black hole physics today, and the TBL models describe
the most general family of dust, i.e. pressureless, collapse solutions.
This result is therefore intriguing, because it shows
that arbitrarily close to the dust black hole model,
we have collapse evolutions with non-zero pressures
that go to a naked singularity final state,
thus proving a certain `instability' of the
OSD black hole formation picture against the introduction of
small pressure perturbations.

Our method consists in `injecting' arbitrarily small but generic
pressure perturbations in a dynamical dust collapse
which was originally going to a black hole final state.
The pressure is chosen in such a way that it remains
small as compared to the energy density during the whole
collapse and the evolution remains close to the corresponding dust
model at all times.
We then examine, when the small
pressures are considered, whether the collapse would
evolve to a black hole or a naked singularity.
Our analysis here shows that in the space
of initial data from which the gravitational collapse
evolves, any arbitrarily small neighborhood of the
OSD model would contain collapse evolutions with
pressure that go to a naked singularity final fate.
While the CCC states that the OSD collapse final
fate is necessarily replicated for any realistic stellar
collapse in nature, the result here shows that an arbitrarily
small pressure perturbation of the OSD model can change
the final outcome of collapse to a naked singularity
and therefore the OSD black hole may be
considered `unstable' in this sense.
Since the pressures within a massive star are very
important physical forces to take into account, we thus
obtain here an important insight into
the stability
of occurrence of black holes as collapse final
states.
It is such a clarification of the structure
of general relativistic collapse, that may provide us
a better understanding of cosmic censorship, and finally resolve
the issue of black hole formation in gravitation
theory.

\section{Small pressure perturbations to dust collapse}

The key feature that characterizes black hole formation in gravitational
collapse is, as collapse evolves, trapped surfaces and
apparent horizon develop at a certain stage within the
collapsing cloud, prior to the epoch of formation
of the spacetime singularity. Then no timelike or null
trajectories can escape from the singularity or its vicinity.
An event horizon then must form hiding the singularity, and
the collapse ends in a black hole final state.
This is the scenario for an homogeneous pressureless
collapse model.
What we show below
is that an arbitrarily small pressure perturbation can radically
alter such a scenario. The trapped surface formation is then
delayed, and this allows the singularity to be naked where
families of non-spacelike curves can escape from the same.
We explicitly identify here such a class of pressure
perturbations, but it is by no means the only class that
can do it. We thus see that the introduction of the slightest
pressure in an otherwise pressure free dust model can
drastically change the final outcome of collapse.

The most general spherically symmetric metric
describing a collapsing matter cloud in comoving coordinates
$(t,r,\theta,\phi)$ is characterized by three free metric
functions $g_{00}=e^{\nu(r,t)}, g_{rr}=e^{\psi(r,t)}$ and the physical radius
of the cloud $R(r,t)=g_{\theta\theta}=g_{\phi\phi}$.
These are related via the Einstein equations with the energy-momentum
tensor, which for perfect fluid matter sources is given by
$T_t^t=-\rho; \; T_r^r=T_\theta^\theta=T_\phi^\phi=p$, where
$\rho$ is matter energy density and $p$ is the pressure
in the cloud. The Einstein equations are then,
\begin{eqnarray}\label{p}
p&=&-\frac{\dot{F}}{R^2\dot{R}} \; , \\ \label{rho}
\rho&=&\frac{F'}{R^2R'} \; ,\\ \label{nu}
\nu'&=&-\frac{p'}{\rho+p} \; ,\\ \label{G}
2\dot{R}'&=&R'\frac{\dot{G}}{G}+\dot{R}\frac{H'}{H} \; ,\\ \label{F}
F&=&R(1-G+H) \; ,
\end{eqnarray}
where the dot and prime represent
derivatives with respect to $t$ and $r$ respectively
and the functions $H$ and $G$ are defined as,
\begin{equation}
    H =e^{-2\nu(r, t)}\dot{R}^2 , \; G=e^{-2\psi(r, t)}R'^2 \; .
\end{equation}
Here $F$ is the Misner-Sharp mass of
the system, representing the amount of matter enclosed
in a radius $r$ at the time $t$.

It is known that since the collapse must evolve
from a regular initial data, close to the center the radial and
tangential pressures must be equal for any general
collapsing matter field.
In fact it can be proven that near the center of the cloud, which is the
region relevant to our purpose, the pressure gradient
must vanish, thus forcing the matter to have a
perfect fluid-like behaviour
\cite{JG}.
Therefore we work here with a perfect fluid matter cloud
as given by the energy-momentum tensor above.

The model has an additional degree of freedom
due to the scale invariance and therefore we can choose
the initial time $t_i$ in such a way that $R(r, t_i)=r$.
We thus introduce the scaling function $v(r,t)$
defined by $R=rv$ with $v(r,t_i)=1$. Collapse is
described by the condition $\dot{v}<0$, and the singularity
is reached at $v=0$, where the density diverges.
The energy density $\rho$ is regular along the central shell
$r=0$ at any time anteceding the singularity, and
that requires the Misner-Sharp mass to have a form
\begin{equation}
    F(r,t)=r^3M(r,v(r,t)) \; ,
\end{equation}
with $M$ being finite at the center.
Also, requiring that the energy
density has no cusps at the center and is hence
a smooth and even function, implies that
$M'(0,v)=0$.
We then have five equations in six
unknowns $\rho$, $p$, $M$, $\nu$, $G$ and $v$ and
the system is closed if an equation of state relating
pressure and density is assumed. In general, however,
it is possible and sometimes even desirable to study
physically valid dynamics satisfying various regularity
and energy conditions, without assuming a priori
any equation of state on which we know little
at very high densities of matter.

We therefore choose a suitable physically motivated
mass function $M$ as the free function. The physical
validity of this choice suggests to consider a mass
profile arbitrarily close to a well-known collapse
scenarios such as the OSD or TBL models. We
therefore deal with an arbitrarily small pressure
perturbation of the TBL model, where by `small'
we mean that the pressure remains much smaller
than the energy density at all times.

Once $M$ is fixed, we can evolve the collapse
using Einstein equations above and a spacetime
singularity develops as collapse final state. Such an
evolution from regular initial data in terms of initial
density and pressure profiles has been studied in detail
and the conditions for the formation of either a black
hole or naked singularity have been worked out
\cite{JG}.
The spacetime singularity, corresponding
to the epoch $v=0$, is written as the time curve $t_s(r)$
which describes the time at which the shell
labeled by $r$ becomes singular
and where $v(t_s(r),r)=0$. In a neighborhood
of the center this is given by,
\begin{equation}\label{ts}
    t_s(r) = t_0 + \chi_2(0)r^2+o(r^3) \; .
\end{equation}
The quantity $\chi_2(0)$ is the tangent to the singularity
curve at the origin, and if it has a positive value
then the singularity turns out to be naked, while
otherwise it would be hidden in a black hole. When it
vanishes, one has to consider the next order
in $r$. Considering collapse of a perfect fluid implies
the Misner-Sharp mass $F$ is in general not conserved,
so we have to match the collapsing cloud with an
exterior generalized Vaidya spacetime, which is always
possible when the pressure of the matter vanishes
at the boundary
\cite{matching}.

The overall behaviour of the collapsing cloud
is determined by the mass function $M(r,t)$, the
evolution $v(r,t)$, and the initial velocity profiles
$b(r)$ for the cloud, which are all not independent
and are governed by Einstein equations.
The special
case of homogeneous perfect fluid is obtained when $M=M(t)$,
$b(r)=k$ and $v=v(t)$, while the inhomogeneous dust case (TBL)
is obtained for $M=M(r)$ (since $p_r=p_\theta$, this implies $p=0$).
Finally the OSD collapse model is obtained when $M=M_0$,
$b(r)=k$ and $v=v(t)$.

In the following, for the sake of simplicity, we will consider
a constant velocity profile given by $b(r)=1$, in analogy with the
marginally bound collapse in TBL models.

We consider an explicit class of perfect fluid
collapse models by introducing a small pressure to the TBL scenario,
to solve Einstein equations in a neighborhood
of the center. To perturb the TBL model with a small
pressure, we must allow in general $v=v(r,t)$,
rather than $v=v(t)$ only and therefore the simultaneous collapse,
which always ends in a black hole, does not happen.
We take the mass profile to be
\begin{equation}
    M=M_0+M_2(v)r^2 \; ,
\end{equation}
where
$M_0$ is a constant. The pressure perturbation is
small when $M_0 \gg \mid M_2 \mid$ at all times.
We immediately see that setting $M_2=C$ reduces
the model to inhomogeneous dust, and $M_2=0$
further gives the Oppenheimer-Snyder-Datt homogeneous
dust case. To start the collapse from an
inhomogeneous dust cloud, triggering the pressure
perturbation at a later stage, we take,
\begin{equation}
    M_2(v)=C+\epsilon(v) \; .
\end{equation}
The initial condition $M_2(1)=C$ implies
$\epsilon(1)=0$, with $v\in[0,1]$.
For the present choice of the mass profile, the
density and pressure, as given by equations \eqref{p}
and \eqref{rho} are,
\begin{equation}
p=-\frac{\epsilon_{,v}}{v^2}r^2 \; , \; \rho= \rho_{TBL}-p+
\frac{5\epsilon-\epsilon_{,v}v}{v^2(v+rv')}r^2 \; ,
\end{equation}
which are fully determined once the function $v$
is derived from the Einstein equations.

We now need to determine the behavior of the
singularity curve $t_s(r)$, and hence to evaluate the
quantity $\chi_2(0)$
\cite{JG}.
To do this, given the choice of $M$ above, one can
integrate \eqref{nu} in the vicinity of $r=0$.
Once $\nu$ is obtained as a function of $M$ and $v$
we can further integrate \eqref{G}. The function
$G$ turns out to be dependent on $M$, $v$ and the
integration function $b(r)$, which, as said before, is related to the
velocity of the collapsing shells. Regularity imposes
some constraints that are easily satisfied in this
case. Equation \eqref{F} then provides the differential
equation for $\dot{v}$ that constitutes the true equation
of motion for the system. Integrating it to obtain
$v(r,t)$ thus solves the set of Einstein equations.
On the other hand, since $v$ is monotonically decreasing
in time, approaching the singularity $v=0$, this equation
can be inverted to give the solution as $t(r,v)$, where
$v$ is now treated as a time coordinate. The singularity
curve is then given by $t_s(r)=t(r,0)$, and can be
expressed as in equation \eqref{ts}.

In the present case, taking $b(r)=1$, we can
evaluate explicitly $\chi_2(0)$ which turns out to be,
\begin{equation}\label{chi2}
    \chi_2(0)=-\frac{1}{2}I_1-\frac{4}{9M_0^2}I_2 \; ,
\end{equation}
with
\begin{eqnarray}
  I_1 &=& \int^1_0(C+Y(v))Z(v)dv \; , \\
  I_2 &=& \int^1_0 W(v)Z(v)dv \; ,
\end{eqnarray}
where we have defined the functions $Y$, $W$ and $Z$ as
\begin{eqnarray}
  Y(v)&=& \left(\epsilon+\frac{2}{3}\epsilon_{,v}v\right) \; , \\
  W(v)&=& \epsilon v (\epsilon+\epsilon_{,v}v) \; , \\
  Z(v)&=& v\left(M_0+\frac{4}{3}\frac{\epsilon v}{M_0}\right)^{-3/2} \; .
\end{eqnarray}

We now analyze the condition for the occurrence
of naked singularity, namely when $\chi_2(0)$ is positive,
and compare the result with the pressureless case.
At first we take $C=0$, by doing so we are perturbing the OSD model.
There are two possibilities for the behaviour
of the function $\epsilon$, namely $\epsilon>0$, which
implies $\epsilon_{,v}<0$ and positive pressure, and
$\epsilon<0$ giving $\epsilon_{,v}>0$ and negative pressure.
We consider here only the case with positive pressures.
Actually, it is easy to see that the negative pressures
more easily favour the occurrence of naked singularities,
therefore it is more useful and physically interesting to
check how positive pressures affect the black hole
formation scenario.

To see the sufficient conditions that must be
required for $\chi_2(0)>0$, we analyze the above two
integrals separately.
Firstly, by evaluating $Z_{,v}$ we see that if we
require,
\begin{equation}
    \text{Condition 1a:} \; M_0^2>\max\{4v\left(\frac{2}{3}
\epsilon+\epsilon_{,v}v\right), v\in[0,1]\} \; ,
\end{equation}
then $Z(v)$ will be monotonically increasing.
We can therefore apply the mean value theorem to
$I_1$, since $Y$ is bounded and integrable for $v\in[0,1]$
and $Z$ is bounded, integrable, monotonic, increasing
and non negative for $v\in[0,1]$. Therefore
$$
I_1=Z(1)\int^1_\eta Y(v)dv \; \text{for some} \; \eta\in[0,1].
$$
Now we analyze the
behaviour of $Y$, noting that $Y(0)=\epsilon(0)>0$
while $Y(1)=\frac{2}{3}\epsilon_{,v}(1)<0$. Therefore,
by Weierstrass theorem, we know that there exist
$\bar{v}\in[0,1]$ such that $Y(\bar{v})=0$ (and $Y<0$
for $v>\bar{v}$, see Fig.1). We shall take $\epsilon(v)$
in such a way that $\bar{v}$ is the only zero of $Y$.
Then if $\eta>\bar{v}$ we have $I_1<0$ and the
first integral in equation \eqref{chi2} will be positive.
If, on the other hand, $\eta<\bar{v}$ then we may
always choose $\epsilon$ in such a way that
\begin{equation}
    \text{Condition 2:} \; \int_{\eta}^{\bar{v}}Y(v)dv<-\int_{\bar{v}}^1Y(v)dv \; ,
\end{equation}
therefore obtaining again $I_1<0$ (see Fig.1).

We turn now our attention to the second integral. We
see immediately that it is typically negligible with
respect to the first one, since it is multiplied by a proportionality
factor $\frac{4}{9M_0^2}$. Nevertheless explicitly requiring
it to be small implies that we choose $M_0$ suitably.
We see that to have $\chi_2>0$ we must have
$$
I_1+\frac{8}{9M_0^2}I_2<0 \; .
$$
Again applying the mean value theorem we see that
$$
I_1+\frac{8}{9M_0^2}I_2=Z(1)\int_\eta^1Y(v)dv+\frac{8}{9M_0^2}W(\omega)Z(1) \; ,
$$
for some $\omega\in[0,1]$. Then $\chi_2$ will be positive
if we choose $M_0$ such that
\begin{equation}
    \text{Condition 1b:} \; M_0^2>-\frac{8}{9}
\frac{W(\omega)}{\int_{\eta}^1Y(v)dv} \;.
\end{equation}
If we choose $\epsilon$ to be at least a cubic
function in $v$ of the type $\epsilon=av^3+bv^2+cv+d$,
it is always possible to choose the four parameters
$a, b, c, d$ in order to fulfill $\epsilon(1)=0$,
$\epsilon_{,v}$ negative and condition 2 (which implies
the values of $\bar{v}$ and $\min\{Y(v)\}$).
Then $M_0$ is chosen in order to satisfy the most stringent
between condition 1a and condition 1b.

\begin{figure}[hh]
\includegraphics[scale=1]{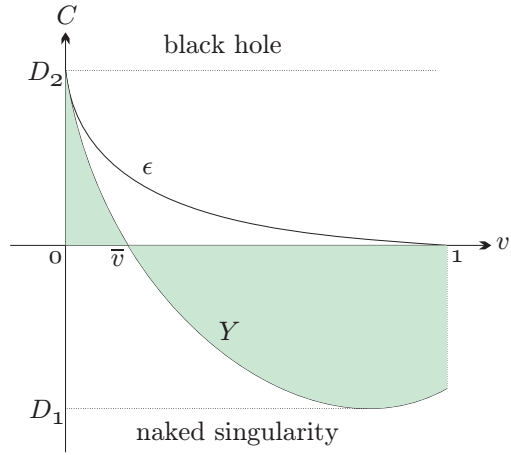}
\caption{Illustrative plot of $\epsilon(v)$ and $Y(v)$ in
the case of naked singularity formation with positive pressures.}
\label{fig}
\end{figure}

We therefore see that it is always possible to
find a suitable positive pressure perturbation of the
OSD collapse model that, however small, will cause the
collapse to end in a naked singularity. Furthermore, it
can be seen that a perturbation with negative pressure
can uncover the OSD black hole in the same manner.
We emphasize that the above consideration is only
to typically illustrate what is possible in collapse.
The above conditions are sufficient for naked singularity
formation but not necessary. In fact, the above example
and class given indicates that there might well be
other classes of perturbations of the OSD model with
small perfect fluid pressures, still leading to a
naked singularity, and a more detailed discussion
will be given elsewhere.

We note that similar considerations also hold
for the pressure perturbations of an inhomogeneous
dust TBL model. The structure is the same as above, and
analogous conclusions hold again, a naked singularity
results when we have $\chi_2>0$. For the sake of clarity
we can see that neglecting the second integral in
equation \eqref{chi2} (which we have shown small
compared to the first one), this is certainly
the case whenever $C$ is chosen such that
\begin{equation}\
    C<D_1=\min\{Y(v), v\in[0,1]\}<0 \; .
\end{equation}
On the other hand, values of $C$ such that
\begin{equation}
    C>D_2=\max\{Y(v), v \in[0,1]\}>0 \; ,
\end{equation}
will lead to the formation of a black hole. For $C\in [D_1,D_2]$
the explicit form of $\epsilon(v)$ is what determines
the sign of $\chi_2$ (see Fig.\ref{fig}). Again,
by putting $\epsilon=0$ we recover the TBL inhomogeneous
dust model, where positive (negative) $C$ leads to black hole
(naked singularity). It is not difficult to see that a
similar reasoning, as given above for the OSD model,
applies here. In fact, in this case from equation \eqref{chi2} we have
\begin{equation}
    \chi_2=\chi_{2|C=0}-\frac{1}{2}\int_0^1CZ(v)dv \; ,
\end{equation}
with $\chi_{2|C=0}$ given by the case $C=0$ studied above and
$\int^1_0 CZ(v)dv$ positive whenever $C>0$, and negative otherwise.
Therefore once we evaluate $\chi_{2|C=0}$ as from the procedure discussed
above we easily see that all those values of $C$ such that
$\frac{1}{2}\int_0^1CZ(v)dv < \chi_{2|C=0}$
will lead to a naked
singularity. It is immediate to see that whenever $\chi_{2|C=0}$
is positive this will include positive values of $C$ that for
the corresponding TBL model were leading to a black hole.
For certain choices of $\epsilon$ we therefore have
models where positive values of $C$ lead to the formation
of a naked singularity, even when the TBL collapse
went to a black hole. Vice versa, for other choices
of $\epsilon$ we can have models where negative
values of $C$ lead to the formation of a black hole,
whereas the corresponding TBL case was leading
collapse to a naked singularity.

We thus see that the structure of Einstein equations
describing gravitational collapse is indeed very rich
and complex. As we see above, both black hole and naked
singularity outcomes are possible, in a general and
generic manner in the sense described above, evolving
from a regular initial data. Of course, in order to
be able to check if the black hole that occurs as the
endstate of the OSD dust collapse is a generic result,
we need an unambiguous definition of genericity,
which we do not have today.

In other words, if we call $\mathcal{G}$ the
set of physically valid initial data for collapse
of a perfect fluid,
this will be divided into the two possible outcomes
as $\mathcal{G}=\mathcal{G}^{BH}\bigcup\mathcal{G}^{NS}$
and every point $I\in\mathcal{G}$ can be characterized
by $I=\{M(r), p(r), b(r)\}$ (so that the OSD initial
configuration is given by $I_{OSD}=\{M_0, 0,
k\}\in\mathcal{G}^{BH}$).
We have therefore shown that for every neighborhood
$\mathcal{U}(I_{OSD})\subset \mathcal{G}$, however small,
there exist (physically valid) pressure profiles
with initial data $I\in\mathcal{U}$ such that
$I\in\mathcal{G}^{NS}$, and similarly for the TBL model.
In this sense we say that the dust collapse model
leading to a black hole is not `stable' under the introduction
of small pressure perturbations.
Considering the mass and pressure profiles to be expandable
in a neighborhood of the center we can move from the space of functions
in which $M$ and $p$ are defined to the space of the relevant parameters
to determine the outcome of collapse. This can give some
further insights in terms of such parameters.

This adds to the known results on the OSD model,
namely that the introduction of inhomogeneities
or a suitable tangential pressure perturbation can
lead to naked singularity formation
\cite{MJ}.
In this sense, we have a strong indication that
the OSD dust model is `unstable' in that
the initial configurations for its endstates lie on
the critical surface separating the two possible
outcomes of collapse discussed above.

\section{Concluding remarks}

We have shown that the introduction of a suitable, though general,
arbitrarily small, perfect fluid pressure perturbation in the well
known OSD collapse model can change
the final outcome from a black hole to a naked singularity.
The pressure, containing only quadratic terms, is chosen in such
a way as to satisfy all usual physical requirements such as
energy conditions and regularity of initial data.

The model presented above lacks
a constitutive relation between the pressure and the energy density.
Once such an equation of state is introduced the system becomes
closed and we do not have the freedom to chose the matter profile
at will. Nevertheless, assuming an equation of state is not
likely to change substantially the picture since the above
model is arbitrarily close to the well known dust models,
and it can be expected to be also arbitrarily close to the
solution in the presence an equation of state for a suitable
choice of the constitutive relation. Further investigation
into perfect fluid collapse with an equation of state is currently
in progress. In this case we still expect the initial
configurations in the pressureless case to act as a critical
surface separating the space of configurations leading to
naked singularity from the one leading to black hole.

The important question would be whether naked
singularities can actually occur in the observable universe
and if so, if they bear signatures in any way different from the
black holes. Indeed, if singularities signal a breakdown
of classical gravity when very high densities are reached
in very small volumes, then these models show that we must
consider the possibility that the region of spacetime
dominated by quantum gravity can affect outside observers
and interact with the rest of the universe.

One should then ask what kind of observational signature
these objects bear, if any, and whether such phenomena can
possibly be observed. The relevant point is if any measurable
amount of energy can come out of such ultra-strong gravity
regions and,
if so, in what form. We know that non rotating spherically
symmetric collapse, settling to a Schwarzschild black hole,
cannot emit gravitational waves, still particles and
photons can in principle escape the ultra-dense region,
thus carrying with them some of the energy
\cite{accel}.
Nevertheless,
a more accurate description of the phenomenon must take into
account rotation and some constitutive relation for the
matter model. At present this is attainable only within the
field of numerical relativity and we hope that future research
will bring some more light on the nature of the final
stages of collapse, since we believe that the next frontier
of black hole physics will lie in further
investigations on the same.

\end{document}